\documentclass[11pt,twoside]{article}

\usepackage{asp2006}
\usepackage{epsf}
\usepackage{psfig}
\usepackage{lscape}

\begin{document}
\title{Why Are Ring Galaxies Interesting?}   
\author{James L. Higdon\altaffilmark{1} and Sarah J. U. Higdon\altaffilmark{1}}   

\altaffiltext{1}{Georgia Southern University, Department of Physics, Statesboro, GA 30458}    

\begin{abstract}Compared with ordinary spirals, the ISM in ring galaxies experiences
markedly different physical conditions and evolution. As a result, ring galaxies
provide interesting perspectives on the triggering/quenching of large scale star
formation and the destructive effects of massive stars on molecular cloud complexes. We use
high resolution radio, sub-mm, infrared, and optical data to investigate the role of
gravitational stability in star formation regulation, factors influencing the ISM's 
molecular fraction, and evidence of peculiar star formation laws and efficiencies in
two highly evolved ring galaxies: Cartwheel and the Lindsay-Shapley ring.
\end{abstract}

\section{Introduction}
Ring galaxies are dramatic examples of galaxy transformation caused by a remarkably
simple interaction. Observations and numerical models (\citealp{lyndstoomre})
argue persuasively that these objects are formed by the near central passage 
of a companion through a spiral along the rotation axis. The brief additional gravitational force 
induces epicyclic motions throughout the disk, which act to form radially
propagating orbit-crowded rings of gas and stars. The concentration of the ISM into 
the expanding ring (at the disk's expense) is nearly total and can last for $\approx400$ Myrs. 
It is this radical rearrangement of the spiral's ISM that is responsible for their interesting
star forming properties.

\section{Star Formation Rates in Ring Galaxies}

Star formation rates ($SFR$) in ring galaxies are typically $\approx5$ M$_{\odot}$ yr$^{-1}$,
i.e., somewhat enhanced over large spirals but far below the $SFR$s inferred in
LIRGs (cf. \citealp{app_struck}; \citealp{higdon95}; \citealp{higdon_wallin};
\citealp{sandersmirabel}). However,
the {\em distribution} of star formation is unique, being completely restricted to 
the expanding rings while simultaneously extinguished over the interior disk. 
Both effects are evident in the Lindsay-Shapley ring ({\em L-S}, hereafter) and Cartwheel, 
shown in Figure 1. A weak nuclear source is responsible for $\la5\%$ of the star formation 
in both.
\begin{figure}[!ht]
   \begin{center}
 	\scalebox{0.45}{\rotatebox{-90} {\includegraphics{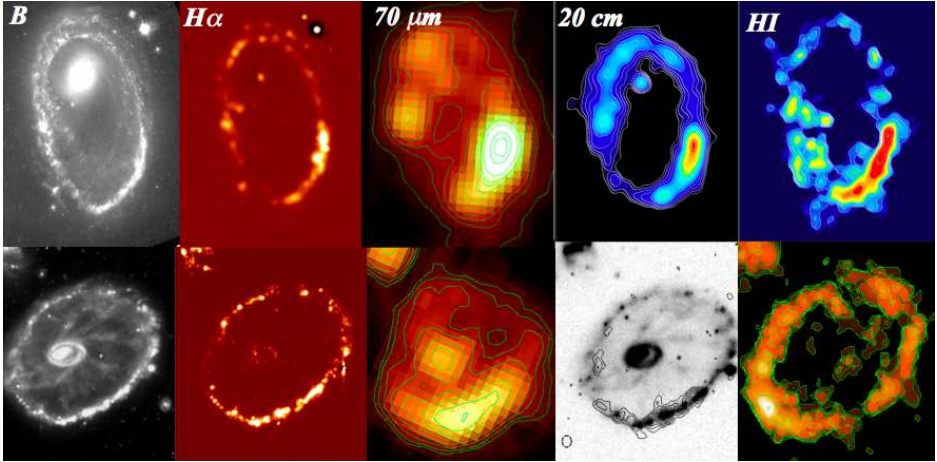}}}
   \end{center}
\caption[]{A polychromatic view of two ring galaxies: ({\em top}) the {\em L-S} ring
galaxy ($SFR = 8$ M$_{\odot}$ yr$^{-1}$, $D_{\rm ring} = 42$ kpc) and ({\em bottom})
the Cartwheel ($SFR = 21$ M$_{\odot}$ yr$^{-1}$, $D_{\rm ring} = 40$ kpc). Massive star
formation is restricted to the rings, which dominate emission at H$\alpha$, far-infrared,
and radio continuum. The right-most column shows the neutral atomic ISM to be likewise
confined to the rings, with low density gas filling the interior (\citealp{higdon10}).
}
\end{figure}
The star forming rings are narrow, with slices showing very sharp radial cutoffs in
H$\alpha$ emission. This implies that OB stars remain in the rings for their Main
Sequence lifetimes, which constrains the stellar velocity dispersion of the rings:
$\sigma_{*} < \Delta r_{\rm ring} / \tau_{\rm OB} \approx 45$ km s$^{-1}$. 
The weak line emission that is sometimes found in ring galaxy disks 
is {\em post}-starburst in origin, i.e., arising from aging HII complexes 
powered by A-stars.

\section{The Interstellar Medium in Ring Galaxies}

	\subsection{Atomic Hydrogen}
{\em L-S} and Cartwheel have been mapped in HI (Figure 1). In such large systems
$\approx95\%$ of the atomic ISM is concentrated in the rings, resulting in
high HI surface densities: $\Sigma_{\rm HI} = 30-120$ M$_{\odot}$ pc$^{-2}$. 
At the same time, their interiors are very gas poor, with $\Sigma_{\rm HI} \la 2$ 
M$_{\odot}$  pc$^{-2}$. HI line-widths are typically narrow ($\sigma_{\rm HI} <10$ km s$^{-1}$). 
Kinematic analysis of the HI yields the ring's
expansion speed ($v_{\rm exp}$), and thus an estimate of the ring galaxy's age
($R_{\rm ring}/v_{\rm exp}$). From their measured radii and $v_{\rm exp}$ ($53 \pm 9$ and 
$154 \pm 10$ km s$^{-1}$, respectively), the rings in Cartwheel and {\em L-S} are $\approx250$ 
and $140$ Myrs old (\citealp{higdon96};\citealp{higdon10}). The $SFR$ in ring galaxies 
correlates with their peak $\Sigma_{\rm HI}$, which explains why young systems (e.g., 
NGC~2793 with $age \approx 50$ Myr) have such low $SFR$: they are still organizing 
their ISM into a 
dense ring.

	\subsection{Molecular Gas in the {\em L-S} Ring Galaxy}
Stars form in cold molecular gas, so HI data can only tell part of the story.
Ring galaxies are not very luminous in the rotational transitions of $^{12}$CO,
a fact often attributed to reduced metallicities from snow-plowing outer disk gas into the 
ring (cf. \citealp{horellou95}). However {\em L-S}'s ring possesses $\approx$solar
abundances (\citealp{fma}), which together with its large angular size, made it 
an ideal target for the $SEST$ (\citealp{higdon10}). We observed 16 positions in 
{\em L-S} in  $^{12}$CO(J=1-0) and $^{12}$CO(J=2-1) transitions: 14 on the ring and one 
each centered on the nucleus and enclosed disk. Figure 2 shows CO detections in 9/14
ring positions, defining two molecular arcs in the ring's north and southwest. The latter 
coincides with the galaxy's peak $\Sigma_{\rm HI}$ and $\Sigma_{\rm H\alpha}$. 
\begin{figure}[!ht]
\epsscale{0.70}
\plotone{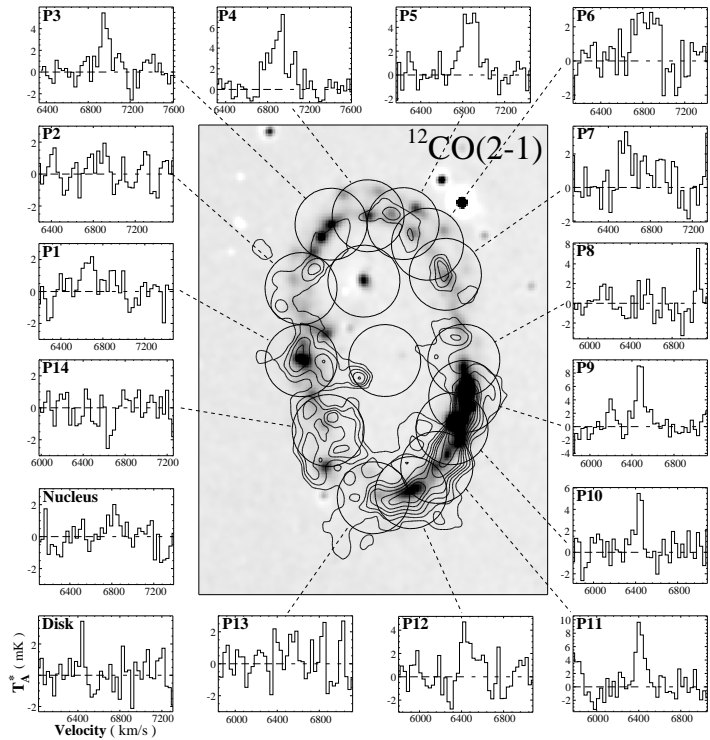}
\caption{H$_2$ in {\em L-S} as traced by $^{12}$CO(J=2-1) emission
using the Swedish ESO Sub-millimeter Telescope ($SEST$). The circles represent the beam
at 230 GHz ($22\arcsec$ FWHM). H$\alpha$ emission is shown in grey-scale (linear stretch), while
contours show HI surface density. Molecular gas is clearly detected at P3-P7 (north) and
P9-P12 (southwest), with broad and complicated line profiles as a rule. The disk and
nucleus are not detected (\citealp{higdon10}).}
\end{figure}
{\em L-S}'s ring is dominated by {\em atomic} rather than molecular gas. For a Galactic 
I$_{\rm CO}$-N$_{\rm H_{2}}$, we find M$_{\rm H_{2}}$/M$_{\rm HI} = 0.06 \pm 0.01$. 
Astonishingly, a typical dwarf galaxy has nearly as much H$_2$ as {\em L-S}'s ring 
(\citealp{leroy}). The molecular gas fraction 
($f_{\rm mol} = M_{H_2}/(M_{\rm HI} + M_{\rm H_2})$) varies
considerably around the ring, and is lowest in the ring's southwest quadrant
($f_{\rm mol} \la 0.03$ at P9-P11), where both $\Sigma_{\rm HI}$ and $\Sigma_{\rm H\alpha}$ peak.

The $^{12}$CO and HI line profiles in {\em L-S}'s ring can be extremely broad 
($\sigma_{\rm gas} = 250-400$ km s$^{-1}$), with multiple velocity components or 
broad tails evident. It is not clear if this represents out-of-plane gas motions or caustics, 
though preliminary numerical models suggest the latter (J. Wallin, private communication).

\section{Star Formation Processes in the Ring}

	\subsection{The Role of Gravitational Instabilities}
The onset of robust star formation in gas disks can be described
in terms of local gravitational stability parameters, e.g.,
$Q_{\rm gas} = \sigma_{\rm gas} \kappa / (\pi  G \Sigma_{\rm gas})$, where
$\sigma_{gas}$ is the gas velocity dispersion and $\kappa$ is the disk's epicyclic
frequency (\citealp{quirk}). Regions of the disk where $Q_{\rm gas} < 1$ are Jeans unstable and 
prone to collapse, leading to the formation of molecular cloud complexes and eventually stars. 
The ``bead on a string'' morphology evident in H$\alpha$ (Figure 1) suggests that the rings 
are gravitationally unstable. Are they?

The Cartwheel's ring is, even ignoring its (unknown) molecular component.
However, $Q_{gas} > 1$ essentially everywhere in {\em L-S}'a ring (Figure 3), due
to the large $\sigma_{\rm gas}$. This ignores, however, the
stellar component's contribution to $Q$. Using IRAC 4.5 $\mu$m data we find that the stellar
mass surface density ($\Sigma_*$) everywhere exceeds that of gas (i.e., $\Sigma_{*} > 
\Sigma_{\rm HI} + \Sigma_{\rm H_2}$). A gravitational stability parameter combining stars and
gas can be written $Q_{\rm tot} = {{\kappa}\over{\pi G}} 
( {{\Sigma_{\rm gas}}\over{\sigma_{\rm gas}}} + {{\Sigma_*}\over{\sigma_*}}  )^{-1}$ 
(\citealp{wang_silk}), where the ring's radial H$\alpha$ profile constrains 
$\sigma_* \la 45$ km s$^{-1}$. Figure 3 shows that when the stellar component is included, 
$Q_{\rm tot} < 1$ and {\em L-S}'s ring is everywhere Jeans unstable. 
\begin{figure}[!ht]
   \begin{center}
 	\scalebox{0.55}{\rotatebox{-90} {\includegraphics{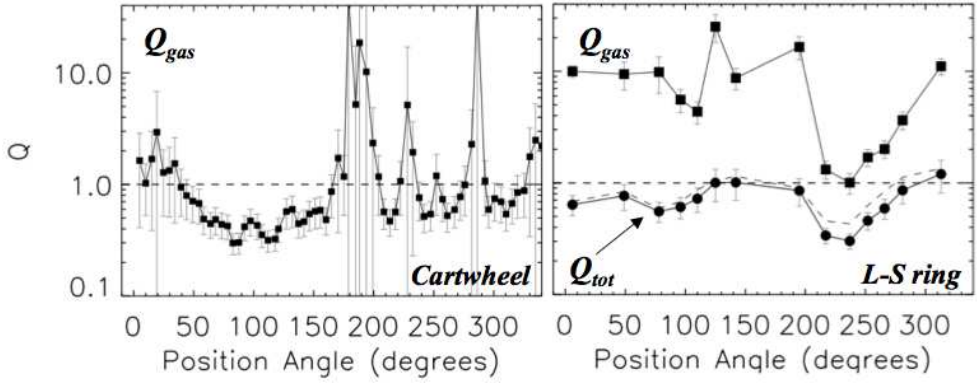}}}
   \end{center}
\caption[]{Azimuthal variations in $Q$ for Cartwheel ({\em left}) and {\em L-S} ({\em right}).
The Cartwheel's HI ring is largely sub-critical ($Q_{\rm gas} < 1$), but not {\em L-S}
even after adding molecular gas. Only when stellar mass is included does the ring become
gravitationally unstable (i.e., $Q_{\rm tot} < 1$) everywhere.}
\end{figure}
Conversely, the interior disks in both satisfy $Q_{\rm tot} > 1$, i.e.,
these regions are stable against the growth of gravitational instabilities and
star formation is effectively quenched.

	\subsection{Evidence for Peculiar Star Formation Laws}
A strong correlation exists between $SFR/area$ ($\Sigma_{\rm SFR}$)
and the surface density of cold gas in galaxies, i.e., the ``Schmidt Law'',
written $\Sigma_{\rm SFR} = \beta \Sigma_{\rm gas}^{\rm N}$. In M~51 for example,
H$_2$ and $SFR/area$ obey $\Sigma_{\rm SFR} \propto \Sigma_{\rm H_{2}}^{1.37 \pm 0.03}$
(Figure 4). HI is uncorrelated with $\Sigma_{\rm SFR}$, implying that it
is a photo-dissociation product and not directly involved in the star formation process.
We show star formation laws derived for the Cartwheel and {\em L-S} in the same figure.
Both are peculiar. Unlike M~51, atomic gas in the Cartwheel correlates with $\Sigma_{\rm SFR}$
in most of the ring, though with a small $N$. The exponent becomes negative 
(i.e., {\em anti}-correlated) where $\Sigma_{\rm SFR}$ peaks. Would this peculiarity
disappear if the molecular component were available? Not if {\em L-S}'s ring is any guide:
atomic gas obeys M~51's {\em molecular} Schmidt Law, but its cold molecular ISM is 
{\em uncorrelated} with $\Sigma_{\rm SFR}$, which is completely the opposite from M~51. 
How can H$_2$ be so apparently disconnected from star formation?
\begin{figure}[!ht]
   \begin{center}
 	\scalebox{0.33}{\rotatebox{-90} {\includegraphics{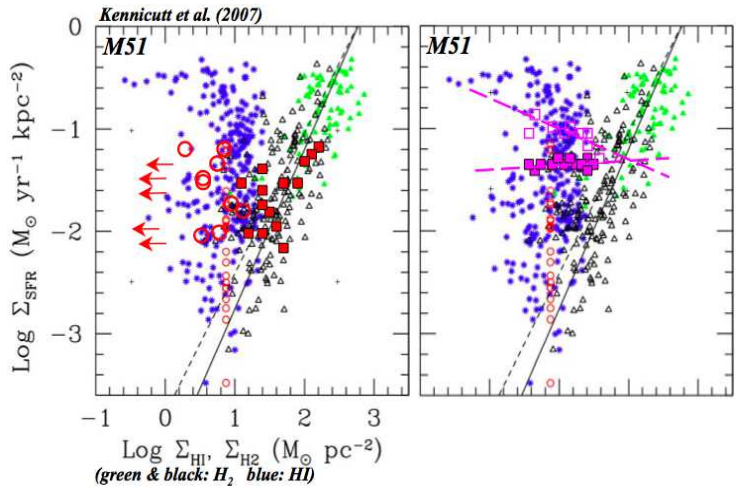}}}
   \end{center}
\caption[]{Star formation laws in the rings of {\em L-S} ({\em left}) and Cartwheel ({\em right})
relative to M~51 (\citealp{kennicutt07}). H$_2$ in M~51 (black \& green triangles) obeys a
Schmidt Law, but HI (blue) is uncorrelated with $\Sigma_{\rm SFR}$. In {\em L-S} the opposite
is true: HI (red squares) obeys a Schmidt Law but H$_2$ is uncorrelated (red circles
\& arrows). In Cartwheel atomic gas can be correlated (filled purple squares) or 
{\em anti}-correlated (empty squares) with $SFR/area$.}
\end{figure}

	\subsection{Enhanced Star Formation Efficiencies}
Star formation efficiency ($SFE$) is the yield of massive stars 
per unit H$_2$ mass. \citet{young96} find nearly constant $SFE$ 
($\equiv \log(\rm L_{\rm H\alpha}/M_{\rm H_2}$)) from
S0 to Scd ($\approx -1.8$ L$_{\odot}$/M$_{\odot}$). Later types show
higher $SFE$, peaking at $-0.8$ L$_{\odot}$/M$_{\odot}$ for Irr. Detecting
molecular gas in {\em L-S}'s ring allows the first estimate of $SFE$ in a
ring galaxy. We find $SFE = -0.7 \pm 0.1$ L$_{\odot}$/M$_{\odot}$, i.e., similar 
to an Irr and an order of magnitude higher than the (presumably $\sim$Sa) progenitor.
This result depends, of course, on our ability to reliably measure H$_2$ in the
ring using $^{12}$CO emission.

\section{Why is the Molecular Gas Fraction So Low?}
The rings are gas rich but seemingly H$_2$ poor. Since HI rapidly converts into H$_2$,
the low $f_{\rm mol}$ cannot signal the consumption of the molecular gas reservoir. Nor
can metallicity effects by themselves be responsible, at least in {\em L-S}. The gas phase 
pressure ($P_{\rm ISM}$) might be a factor, as it directly affects the HI to H$_{2}$ conversion 
rate. \citet{bruce93} finds $f_{\rm mol} 
\approx (P_{\rm ISM}/P_{\odot})^{2.2} (\chi/\chi_{\odot})^{-1}$, 
where $\chi$ is the ambient UV-field. For {\em L-S}'s ring we estimated $\chi$
with Spitzer and GALEX images, and $P_{\rm ISM} \approx (\pi G/2)(\Sigma_{\rm gas}^{2}
+ {{\sigma_{\rm gas}}\over{\sigma_*}}\Sigma_{\rm gas}\Sigma_{*})$, with the ring's
stellar surface mass density derived with IRAC 4.5 $\mu$m data. We find very high gas phase 
pressures, with $P_{\rm ISM}/P_{\rm ISM,local} \approx 30-400$, leading us to expect 
$f_{\rm mol} \approx 1$ everywhere. {\em L-S}'s ring should be dominated by 
molecular gas. Why isn't it? 

We used photo-dissociation models in \citet{allen04} to estimate average
gas volume densities ($n$) in {\em L-S}'s ring given its $\Sigma_{\rm HI}$ and UV-field. 
In the northern half of the ring,
$n = 100-300$ cm$^{-3}$, implying an ISM dominated by the {\em Cold Neutral
Medium} ({\em CNM}, $T = 50-100$ K), i.e., the precursor of cold molecular clouds. In the southwest, 
where $\Sigma_{\rm SFR}$ and $\Sigma_{\rm HI}$ are both much higher, the models give
$n \approx 2$ cm$^{-3}$, which taken at face value, points to an ISM dominated by the
{\em Warm Neutral Medium} ({\em WNM}, $T \approx 7000$ K). How can you form stars out of this?

We believe the answer lies in fundamental differences in the environments of rings
and spiral arms. Consider that molecular clouds spend $\approx20$ Myrs in the arms of grand
design spirals like M~51, whereas the ISM is confined in rings, equally as dense and actively
forming stars, for $\approx200$ Myr. While molecular cloud growth is enhanced 
in the high $\Sigma_{\rm gas}$ rings, the destructive effects of SNe and OB stars 
are also amplified. A dominant {\em WNM} might be expected as the molecular clouds 
become fragmented and ``over-cooked'' by shocks and sustained UV-fields.
CO might in this case retreat to the inner-most cloud cores resulting in weak $I_{\rm CO}$
and underestimates of $\Sigma_{\rm H_{2}}$. This might explain the peculiar Schmidt Laws and
enhanced $SFE$.  At the same time, higher cloud collision rates might favor the formation
of unusually large molecular cloud complexes and more efficient star formation. More
work remains though results from Cartwheel and {\em L-S} are intriguing.

\section{Summary and Future Prospects}
The distribution and gravitational stability of a ring galaxy's ISM changes
dramatically as it evolves.  Further, the long confinement in the dense star forming
ring is expected result in a fragmented and largely atomic ISM, which may explain
the peculiar star formation laws, efficiencies, and $f_{\rm mol}$ observed in
the two largest and most evolved ring galaxies, Cartwheel and {\em L-S}. Future 
progress will require sensitive and high resolution assays of the molecular
ISM in these and other ring galaxies, which will be possible with ALMA and CARMA.

%%% THE BIBLIOGRAPHY


\begin{thebibliography}{}

\bibitem[Appleton \& Struck(1987)]{app_struck}
Appleton, P. \& Struck, C. J. 1987, \apj, 312, 103

\bibitem[Allen et al.(2004)]{allen04}
Allen, R. J., Heaton, H., \& Kaufmann, M. J. 2004, \apj, 608, 314

\bibitem[Elmegreen(1993)]{bruce93}
Elmegreen, B. E.  1993, \apj, 411, 170

\bibitem[Few et al.(1982)]{fma}
Few, J. M., Madore, B., \& Arp, H. 1982, \mnras, 199, 633

\bibitem[Higdon(1995)]{higdon95}
Higdon, J. L. 1995, \apj, 455, 524

\bibitem[Higdon(1996)]{higdon96}
Higdon, J. L. 1996, \apj, 467, 241

\bibitem[Higdon \& Wallin(1997)]{higdon_wallin}
Higdon, J. L. \& Wallin, J. F. 1997, \apj, 474, 686

\bibitem[Higdon et al.(2010)]{higdon10}
Higdon, J. L., Higdon, S. J. U., Rand, R. J., \& Nord, M. 2010, submitted to \apj

\bibitem[Horrelou et al.(1995)]{horellou95}
Horrelou, C., Casoli, F., Combes, F., \& Dupraz, C.  1995, \aap, 298, 743

\bibitem[Kennicutt et al.(2007)]{kennicutt07}
Kennicutt, R. C. et al. 2007, \apj, 671, 333

\bibitem[Leroy et al.(2005)]{leroy}
Leroy, A. et al. 2005, \apj, 625, 763

\bibitem[Lynds \& Toomre(1976)]{lyndstoomre}
Lynds, R. \& Toomre, A. 1976, \apj, 209, 382

\bibitem[Quirk(1972)]{quirk}
Quirk, W. 1972, \apj, 176, L9

\bibitem[Sanders \& Mirabel(1996)]{sandersmirabel}
Sanders, D. \& Mirabel, I. 1996, \araa, 34, 749

\bibitem[Wang \& Silk(1994)]{wang_silk}
Wang, B. \& Silk, J. 1994, \apj, 427, 759

\bibitem[Young et al.(1996)]{young96}
Young, J. et al. 1996, \aj, 112, 1903

\end{thebibliography}
\end{document}